\documentclass[superscriptaddress,prl,twocolumn,nofootinbib,showpacs,preprintnumbers]{revtex4}
\usepackage{graphicx,epsfig}
\begin{document}

\title{String Inspired Singlet Extensions of the Minimal
Supersymmetric Standard Model}

\author{Tianjun Li}

\affiliation{George P. and Cynthia W. Mitchell Institute for
Fundamental Physics, Texas A$\&$M University, College Station, TX
77843, USA }

\affiliation{ Institute of Theoretical Physics, Chinese Academy of
Sciences, Beijing 100080, China}



\begin{abstract}

The only allowed Higgs superpotential term at stringy tree level in
the string derived Singlet Extensions of the Minimal Supersymmetric
Standard Model (SEMSSM) is $h S H_d H_u$, which leads to an
additional global $U(1)$ symmetry in the Higgs potential. We propose
the string inspired SEMSSM where the global $U(1)$ symmetry is
broken by the additional superpotential terms or supersymmetry
breaking soft terms that can be obtained naturally due to the
instanton effects or anomalous $U(1)_A$ gauge symmetry. In these
models, we can solve the $\mu$ problem and the fine-tuning problem
for the lightest CP-even Higgs boson mass in the MSSM, generate the
baryon asymmetry via electroweak baryogenesis, and predict the new
Higgs physics which can be tested at the LHC and ILC.

\end{abstract}


\preprint{MIFP-06-26}

\maketitle


{\bf Introduction~--}~The Minimal Supersymmetric Standard Model
(MSSM) can solve the gauge hierarchy problem elegantly due to
supersymmetry, has neutralino as cold dark matter candidate, and
accommodates the gauge coupling
unification~\cite{Langacker:1990jh,Amaldi:1991cn}.
 So, it is the most natural extension of the Standard Model (SM).
However, there are a few problems within the MSSM. The bilinear
supersymmetric Higgs mass term $\mu H_d H_u$ in the superpotential,
where  $H_d$ and $H_u$ are
 one pair of Higgs doublets, does  not violate supersymmetry and
gauge  symmetry. Then the natural scale for $\mu$ is about Planck
scale but not the TeV scale, which leads to the $\mu$ problem.
Moreover, in order to have the lightest CP-even Higgs boson mass
larger than the low bound 114 GeV from the LEP experiment, there
exists a few percent fine-tuning~\cite{Okada:1990vk}.

To solve the $\mu$ problem, the Next to the Minimal Supersymmetric
Standard Model (NMSSM) was proposed in which a SM singlet $S$ and a
$Z_3$ discrete symmetry are introduced~\cite{Nilles:1982dy}. The
$\mu H_d H_u$ term is forbidden by the $Z_3$ symmetry, and the
superpotential in the NMSSM is
\begin{eqnarray}
W &=& h S H_d H_u + {{\kappa}\over {3!}} S^3 ~,~\, \label{NMSSM-SP}
\end{eqnarray}
where $h$ and $\kappa$ are Yukawa couplings. After $S$ obtains a
vacuum expectation value (VEV), the effective $\mu$ term $\mu_{\rm
eff}=h \langle S \rangle$ is generated. Also, the F-term of $S$ will
give additional contribution to the Higgs quartic coupling, and then
can increase the lightest CP-even Higgs boson mass. In addition, the
lightest CP-even Higgs boson in the NMSSM can have mass around
100~GeV because of its invisible decay~\cite{Dermisek:2005ar}, and
the above fine-tuning problem for its mass can be solved. Moreover,
the observed baryon asymmetry can be generated via electroweak
baryogenesis because there are extra CP violating phases in the
supersymmetry breaking soft parameters and the trilinear soft term
$A_h h S H_d H_u$ can give us strong first order electroweak phase
transition~\cite{Pietroni:1992in}. Similar results hold for the
nearly MSSM~\cite{{nMSSM}} and the $U(1)'$-extended supersymmetric
Standard Models~\cite{UMSSM}. Therefore, the Singlet Extensions of
the Minimal Supersymmetric Standard Model (SEMSSM) is very
interesting from phenomenological point of view.

On the other hand, string theory may be the only known theory which
can correctly describe the quantum gravity. In string models, we may
solve the $\mu$ problem in the MSSM~\cite{Blumenhagen:2006xt}, and
the doublet-triplet splitting problem in the Grand Unified Theories
(GUTs)~\cite{Braun:2005ux,Chen:2006ip}. Thus, how to indirectly test
string models at the Large Hadron Collider (LHC) and the
International Linear Collider (ILC) is a pretty interesting
question.

As we know, in the string model building, the renormalizable terms
in the superpotential, which arise from the Chern-Simmons terms in
the heterotic string compactification~\cite{Li:1997sk} or instanton
effects (triangles formed by the intersections of D6-branes) in Type
IIA intersecting D6-brane models~\cite{Cremades:2003qj}, have the
following generic trilinear form at stringy tree level
\begin{eqnarray}
W &=&  y_{\phi} \phi_1 \phi_2 \phi_3 ~,~\,
\end{eqnarray}
where $y_{\phi}$ is the Yukawa coupling, and $\phi_i$ are different
fields. Thus, the $\mu H_d H_u$ term in the MSSM and the $\kappa
S^3/3!$ term in the NMSSM do not exist at stringy tree level in the
string derived models. And only the first term  $h S H_d H_u$ in the
superpotential in Eq. (\ref{NMSSM-SP}) is allowed where $S$ is a
modulus. With only $h S H_d H_u$ term in the superpotential, we have
two global $U(1)$ symmetries in the Higgs potential in which one of
them is $U(1)_Y$ gauge symmetry. So, there is one global $U(1)$
symmetry, and then there exists one massless Goldstone boson with
$S$, $H_d^0$ and $H_u^0$ mixing components, which is excluded from
the known experiments.

In this letter, we propose the string inspired SEMSSM. The global
$U(1)$ symmetry in the Higgs potential is broken by the additional
 superpotential terms or  supersymmetry breaking soft
terms. The extra superpotential terms can be realized in the string
derived models via instanton effects~\cite{Blumenhagen:2006xt}. With
anomalous $U(1)_A$ gauge symmetry~\cite{Dreiner:2003yr}, we
construct four simple and concrete SEMSSM.  In these models, we can
naturally solve the $\mu$ problem and the fine-tuning problem for
the lightest CP-even Higgs boson mass in the MSSM. We also calculate
the Higgs boson masses, chargino masses and neutralino masses at
tree level, and predict the new Higgs physics which can be tested at
the LHC and ILC. A more detail discussions will be presented
elsewhere~\cite{Tianjun}.

{\bf Model Building~--}~Let us consider the most general SEMSSM. The
generic superpotential is
\begin{eqnarray}
W &=&  h S H_d H_u + \mu H_d H_u + m^2 S +
 {{\mu'}\over {2!}} S^2 +
{{\kappa}\over {3!}} S^3 ~,~\, \label{Superpotential-General}
\end{eqnarray}
where $\mu$, $m^2$, and $\mu'$ are mass parameters. The
corresponding $F$-term scalar potential  is
\begin{eqnarray}
V_F &=& |h H_d H_u +m^2+ \mu' S+{{\kappa}\over {2!}} S^2|^2
\nonumber\\&& +|h S +\mu|^2 |H_u|^2  +|h S +\mu|^2 |H_d|^2 ~.~\,
\end{eqnarray}
And the $D$-term scalar potential is
\begin{eqnarray}
V_D &=& {{G^2}\over 8} \left(|H_u|^2 - |H_d|^2\right)^2~,~\,
\end{eqnarray}
where $G^2=g_Y^{2} +g_2^2$; $g_Y$ and $g_2$ are respectively the
coupling constants for $U(1)_Y$ and $SU(2)_L$. Moreover, we
introduce the supersymmetry breaking soft terms $V^I_{soft}$ and
$V^{II}_{soft}$ as follows
\begin{eqnarray}
V^I_{soft} &=& m_{H_d}^2 |H_d|^2 + m_{H_u}^2 |H_u|^2 + m_S^2
|S|^2~,~
\label{vsoftI} \\
V^{II}_{soft} &=& -\left(A_h h S H_d H_u + B \mu_B H_d H_u
 + A_X m_X^2 S \right.\nonumber\\&& \left.
+ {1\over {2!}} B^{\prime} \mu^{\prime}_B S^2 +{1\over {3!}}
A_{\kappa} \kappa_X S^3 + {\rm H.C.} \right) ~,~\,
 \label{vsoftII}
\end{eqnarray}
where $m^2_{H_d}$,  $m^2_{H_u}$, and $m^2_{S}$ are supersymmetry
breaking soft masses, $A_h$, $B$, $\mu_B$, $A_X$, $m_X^2$, $B'$,
$\mu^{\prime}_B$, and $A_{\kappa}$ are supersymmetry breaking soft
mass parameters, and $\kappa_X$ is the coupling constant. In
addition, if $\mu \not=0$, $m^2\not=0$, $\mu'\not=0$, or
$\kappa\not=0$, we assume $\mu_B=\mu$, $m_X^2=m^2$,
$\mu^{\prime}_B=\mu^{\prime}$, or $\kappa_X=\kappa$, respectively.
However, even if $\mu=0$, $m^2=0$, $\mu'=0$, or $\kappa=0$, we can
show that $\mu_B$, $m_X^2$, $\mu^{\prime}_B$, or $\kappa_X$ might
not be zero in general, so the global $U(1)$ symmetry in the Higgs
potential can be broken by the supersymmetry breaking soft
terms~\cite{Tianjun}.

In the string derived models,  the terms $\mu H_d H_u$, $\mu'
S^2/2!$, and $\kappa S^3/3!$ in superpotential in Eq.
(\ref{Superpotential-General}), which are forbidden at stringy tree
level, might be generated due to the instanton effects. And the
effective $\mu$ is about $M_{\rm string} e^{-A}$ where $M_{\rm
string}$ is the string scale around $10^{17}~{\rm GeV}$. So, the
$\mu$ problem in the MSSM is solved if $A \sim
33$~\cite{Blumenhagen:2006xt}. Similar result holds for $\mu'$.
However, the $\kappa S^3/3!$ term generated from instanton effects
might be highly suppressed.

To construct the string inspired SEMSSM, we consider the models with
an anomalous $U(1)_A$ gauge symmetry~\cite{Dreiner:2003yr}. In
string model building, there generically exists one anomalous $U(1)$
gauge symmetry in the heterotic string model
building~\cite{Dreiner:2003yr} or up to four in the Type II
orientifold model building~\cite{Ibanez:2001nd}. The corresponding
anomalies are cancelled by the (generalized) Green-Schwarz
mechanism~\cite{Green:1984sg}. We introduce a SM singlet field
$\phi$ with $U(1)_A$ charge $-1$. To cancel the Fayet-Iliopoulos
term of $U(1)_A$, we assume that $\phi$ obtains a VEV so that the
$U(1)_A$ D-flatness and supersymmetry can be preserved.
Interestingly, $\langle \phi \rangle/M_{\rm Pl}$ is about 0.22,
where $M_{\rm Pl}$ is the Planck scale~\cite{Dreiner:2003yr}.
Moreover, to break the supersymmetry, we introduce a hidden sector
superfield $Z$ whose F component acquires a VEV around $10^{21}~{\rm
GeV}^2$.

We assume that the $U(1)_A$ charges for $S$ and $Z$ are $n+p/q$ and
$m+p'/q'$, respectively, where $m$ and $n$ are integers, ($p$, $q$)
and ($p'$ and $q'$) are relatively prime positive integers, or $p/q$
or $p'/q'$ is zero. To have the $h S H_d H_u$ term in
superpotential, we also assume that the total $U(1)_A$ charges for
$H_d$ and $H_u$ are $-n-p/q$, but we do not give the explicit
charges for $H_d$ and $H_u$ which are irrelevant to our discussions.
Moreover, if $m+p'/q'$ is non-zero, the gaugino masses can not be
generated via F-terms $ZW^{\alpha} W_{\alpha}/M_{\rm Pl}$. To have
the gaugino masses, we can introduce another $U(1)_A$-uncharged
hidden-sector superfield $Z'$ whose F component acquires a VEV. In
fact, in the string model building, both  dilaton and moduli fields
can break the supersymmetry due to their F-component VEVs. In
addition, the supersymmetry breaking soft mass terms in $V^I_{soft}$
can be generated via D-term operators
\begin{eqnarray}
\int d^4x d^2\theta d^2\overline{\theta} {{\overline{Z} Z} \over
{M_{\rm Pl}^2}} \left(  |S|^2+  |H_d|^2+  |H_u|^2 \right)~,~\,
\end{eqnarray}
where for simplicity we neglect the coefficients of these operators
in such kind of discussions in this paper. The first term $A_h h S
H_d H_u$ in $V^{II}_{soft}$ can be generated via the following
F-term operator
\begin{eqnarray}
\int d^4x d^2\theta  {{Z ~({\rm or} ~Z')}\over {M_{\rm Pl}}} h S H_d
H_u + {\rm H. C.}~.~
\end{eqnarray}

 {\em Model A~--}~We choose the following $U(1)_A$ charges for $S$
and $Z$
\begin{eqnarray}
m+n~=~47, ~~p/q~=~1/5 ~,~~p'/q'=4/5~.~\,
\end{eqnarray}
Then the $U(1)_A$ allowed renormalizable superpotential is
\begin{eqnarray}
W &=&  h S H_d H_u ~.~\, \label{Superpotential-Model-A}
\end{eqnarray}
 The additional supersymmetry breaking soft
term $V^{II}_{soft}$ can be generated via the following operator
(the other operators are forbidden by $U(1)_A$ or negligible)
\begin{eqnarray}
\int d^4x d^2\theta M_{\rm string} Z S \left({{\phi}\over {M_{\rm
Pl}}}\right)^{48}  + {\rm H. C.}~.~\,
\end{eqnarray}
So, we have
\begin{eqnarray}
V^{II}_{soft} &=& -\left(A_h h S H_d H_u
 + A_X m_X^2 S  + {\rm H.C.} \right) ~,~\,
 \label{vsoftII-MA}
\end{eqnarray}
where $A_h \sim A_X \sim  10^2~{\rm GeV}$, and $m_X^2 \sim
10^{4-6}~{\rm GeV}^2$. Interestingly, the global $U(1)$ symmetry in
the Higgs potential is indeed broken by the supersymmetry breaking
soft term $A_X m_X^2 S$.

{\em Model B~--}~We choose the following $U(1)_A$ charges for $S$
and $Z$
\begin{eqnarray}
n=-22, ~~p/q~=~0 ~,~~m=0,~~p'/q'=0~.~\,
\end{eqnarray}
The additional relevant F-term and D-term operators are
\begin{eqnarray}
\int d^4x d^2\theta \left(M_{\rm string} H_d H_u + ZH_dH_u \right)
\left({{\phi}\over {M_{\rm Pl}}}\right)^{22} \nonumber
\\
+ \int d^4x d^2\theta d^2{\overline{\theta}} \left(\overline{Z} S +
{{\overline{Z} ZS}\over {M_{\rm Pl}}} \right)
\left({{\overline{\phi}}\over {M_{\rm Pl}}}\right)^{22} + {\rm H.
C.}~.~
\end{eqnarray}
 Thus, the superpotential in Model B is
\begin{eqnarray}
W &=&  h S H_d H_u + \mu H_d H_u + m^2 S ~,~\,
\label{Superpotential-Model-B}
\end{eqnarray}
where $\mu \sim 10^{2-3}~{\rm GeV}$ and $m^2 \sim 10^{4-6}~{\rm
GeV}^2$. And the supersymmetry breaking soft terms in
$V^{II}_{soft}$ are
\begin{eqnarray}
V^{II}_{soft} &=& -\left(A_h h S H_d H_u
 +B \mu_B H_d H_u \right.\nonumber\\&& \left.
 + A_X m_X^2 S
 + {\rm H.C.} \right),\,
 \label{vsoftII-MB}
\end{eqnarray}
where $A_h \sim B \sim \mu_B \sim A_X  \sim 10^{2-3}~{\rm GeV}$, and
$m_X^2 \sim 10^{4-6}~{\rm GeV}^2$.

 {\em Model C~--}~In Model B, we
consider the gauge mediated supersymmetry breaking scenario where
the VEV of F component of $Z$ can be about $10^{10}~{\rm GeV}^2$.
And then the tadpole term $m^2 S$ in the superpotential can be
neglected. Thus, the superpotential in Model C is
\begin{eqnarray}
W &=&  h S H_d H_u + \mu H_d H_u  ~.~\,
\label{Superpotential-Model-C}
\end{eqnarray}
 And the supersymmetry breaking soft
terms in $V^{II}_{soft}$ are
\begin{eqnarray}
V^{II}_{soft} &=& -\left(A_h h S H_d H_u
 +B \mu_B H_d H_u \right).\,
 \label{vsoftII-MC}
\end{eqnarray}

Model C can also be considered as the string derived model with $h S
H_d H_u$ superpotential term where the extra $\mu H_d H_u$ term
arises from instanton effects~\cite{Blumenhagen:2006xt}.

{\em Model D~--}~We choose the following $U(1)_A$ charges for $S$
and $Z$
\begin{eqnarray}
n=11, ~~p/q~=~1/2 ~,~~m=0,~~p'/q'=0~.~\,
\end{eqnarray}
The additional relevant  F-term  operators are
\begin{eqnarray}
\int d^4x d^2\theta \left(M_{\rm string} S^2 + ZS^2\right)
\left({{\phi}\over {M_{\rm Pl}}}\right)^{23} + {\rm H. C.} ~.~
\end{eqnarray}
So, the superpotential is
\begin{eqnarray}
W &=&  h S H_d H_u + {{\mu'}\over {2!}} S^2~,~\,
\label{Superpotential-Model-D}
\end{eqnarray}
where  $\mu' \sim 10^2~{\rm GeV}$.  And the supersymmetry breaking
soft terms in $V^{II}_{soft}$ are
\begin{eqnarray}
V^{II}_{soft} = -\left(A_h h S H_d H_u
 +  {1\over {2!}} B^{\prime} \mu^{\prime}_B S^2
 + {\rm H.C.} \right),\,
 \label{vsoftII-MD}
\end{eqnarray}
where $A_h \sim B' \sim \mu'_B  \sim 10^{2-3}~{\rm GeV}$.

 Model D can be considered as the string derived model with $h S
H_d H_u$  superpotential term where the extra $\mu' S^2/2!$ term
arises from instanton effects~\cite{Blumenhagen:2006xt}. However,
there exists a $Z_4$ symmetry in Model D, where $H_d$ and $H_u$ have
charge 1, and $S$ has charge 2. To avoid the domain wall problem
after symmetry breaking, we can turn on tiny instanton effects to
break the $Z_4$ symmetry by generating small high-dimensional
operators, and then we can dissolve the domain wall.


\renewcommand{\arraystretch}{1.4}
\begin{table}[t]
\caption{The Higgs VEVs, and the charged, CP-even, and  CP-odd
 Higgs boson masses in GeV at tree level.
\label{tab:Higgs}} \vspace{0.4cm}
\begin{center}
\begin{tabular}{|c|c|c|c|c|c|c|c|c|c|}
\hline Model & $\langle H_d^0 \rangle $ &
 $\langle H_u^0 \rangle $ & $\langle S \rangle $ &
$H^{\pm}$ & $H_1^0$  & $H_2^0$ & $H_3^0$ &
$A_1^0$ & $A_2^0$   \\
\hline A & 119 & 127 & 213
& 205 & 67 & 196 & 210 & 127 & 251 \\
\hline B & 123 & 123 & 188
& 179 & 45 & 184 & 206 & 142 & 214 \\
\hline C & 123 & 123 & 161
& 165 & 66 & 148 & 171 & 31 & 214 \\
\hline D & 120 & 126 & 167
& 176 & 67 & 145 & 181 & 39 & 225 \\
\hline
\end{tabular}
\end{center}
\end{table}

{\bf Phenomenological Consequences~--}~We shall calculate the Higgs
boson masses, the chargino and neutralino  masses at tree level in
our models where we neglect the loop corrections for simplicity. The
input parameters with dimensions of mass or mass-squared are chosen
in arbitrary units. After finding an acceptable minimum they are
rescaled so that $\sqrt{\langle H_d^0 \rangle^2+ \langle H_u^0
\rangle^2} \simeq 174.1$ GeV. For Model A, we choose: $h=0.7$,
$m_{H_d}^2 =-0.1$, $m_{H_u}^2 =-0.2$, $m_{S}^2 = 0.1$, $A_h = 1.0$,
$A_X=0.68$, $m_X^2=0.6$. And the VEVs for the Higgs fields at the
minimum are $\langle H_d^0 \rangle =0.7031$, $\langle H_u^0 \rangle
=0.75$, and $\langle S \rangle =1.2563$. For Model B, we choose:
$h=0.7$, $\mu=-0.2$, $m^2=-0.3$, $m_{H_d}^2 =-0.1$, $m_{H_u}^2
=-0.1$, $m_{S}^2 = 0.1$, $A_h = 0.6$, $B=-0.1$, $\mu_B=-0.2$,
$A_X=-1.9$, $m_X^2=-0.3$. And the VEVs for the Higgs fields are
$\langle H_d^0 \rangle =0.8625$, $\langle H_u^0 \rangle =0.8625$,
and $\langle S \rangle =1.3156$. For Model C, we choose: $h=0.7$,
$\mu=-0.1$, $m_{H_d}^2 =-0.1$, $m_{H_u}^2 =-0.1$, $m_{S}^2 = -0.6$,
$A_h = 2.0$, $B=-0.6$, $\mu_B=-0.1$. And the VEVs for the Higgs
fields are $\langle H_d^0 \rangle =1.5875$, $\langle H_u^0 \rangle
=1.5875$, and $\langle S \rangle =2.075$. For Model D, we choose:
$h=0.7$, $\mu'=-0.3$, $m_{H_d}^2 =-0.1$, $m_{H_u}^2 =-0.4$, $m_{S}^2
= -0.68$, $A_h = 2.0$, $B'=-0.6$, $\mu'_B=-0.3$. And the VEVs for
the Higgs fields are $\langle H_d^0 \rangle =1.6375$, $\langle H_u^0
\rangle =1.7203$, and $\langle S \rangle =2.275$.

We present the Higgs VEVs, the charged Higgs boson ($H^{\pm}$) mass,
CP-even Higgs boson ($H_1^0$, $H_2^0$, and $H_3^0$) masses, and
CP-odd Higgs boson ($A_1^0$ and $A_2^0$) masses in Table
\ref{tab:Higgs}. Interestingly, the couplings of the CP-even Higgs
boson $H_1^0$ and $H_2^0$ with $Z^0$ gauge boson almost vanish, and
only the heaviest CP-even Higgs boson $H_3^0$ can couple to
$Z^0$~\cite{Tianjun}. Thus, the fine-tuning problem for the lightest
CP-even Higgs boson mass in the MSSM from the LEP constraints can be
completely relaxed, and we have new Higgs physics at the LHC and ILC
because $H_3^0$ has mass around 190 GeV. Moreover, to calculate the
chargino and neutralino masses, we choose the positive and negative
gaugino masses $M_1$ and $M_2$ for $U(1)_Y$ and $SU(2)_L$: (1)
$M_1=150$ GeV, and $M_2=300$ GeV; (2) $M_1=-150$ GeV, and $M_2=-300$
GeV. The masses for charginos ($\tilde \chi_1^{\pm}$ and $\tilde
\chi_2^{\pm}$)  and neutralinos ($\tilde \chi_i^{0}$ with $i=1, 2,
..., 5$) are given in Table \ref{tab:char-neut}.

\renewcommand{\arraystretch}{1.4}
\begin{table}[t]
\caption{The chargino and neutralino masses in  GeV.
 \label{tab:char-neut}} \vspace{0.4cm}
\begin{center}
\begin{tabular}{|c|c|c|c|c|c|c|c|c|}
\hline Model & $M_i$ & $\tilde \chi_1^{\pm}$ & $\tilde \chi_2^{\pm}$
 & $\tilde \chi_1^{0}$  & $\tilde \chi_2^{0}$ & $\tilde \chi_3^{0}$  &
  $\tilde \chi_4^{0}$ & $\tilde \chi_5^{0}$  \\
\hline
 A & $> 0$ & 115 & 334 &  68 & 88 & 175 & 217 & 336 \\
\hline
 A & $ < 0$ & 163 & 314 & 68 & 156 & 169 & 217 & 314 \\
\hline
 B & $> 0$ & 75 & 328 & 56 & 81 & 167 & 184 & 330 \\
\hline
 B & $ < 0$ & 118 & 315 & 81 & 125 & 156 & 184 & 316 \\
\hline
 C & $> 0$ & 76 & 329 & 58 & 80 & 167 & 185 & 330 \\
\hline
 C &  $ < 0$ & 120 & 315 & 80 & 127 & 156 & 185 & 316 \\
\hline
 D & $> 0$ & 87 & 330 & 60 & 68 & 169 & 200 & 331 \\
\hline
 D & $ < 0$ & 132 & 315 & 61 & 138 & 156 & 200 & 315\\
\hline
\end{tabular}
\end{center}
\end{table}

{\bf Conclusions~--}~In the string derived SEMSSM, $h S H_d H_u$ is
the only allowed superpotential term at stringy tree level. Then
there exist an additional global $U(1)$ symmetry in the Higgs
potential, and the axion problem. We propose the string inspired
SEMSSM in which the global $U(1)$ symmetry is broken by the
additional superpotential terms or  supersymmetry breaking soft
terms. The extra superpotential terms can be obtained via instanton
effects in the string derived models. With anomalous $U(1)_A$ gauge
symmetry, we present four simple and concrete SEMSSM. In these
models, we can naturally solve the $\mu$ problem and the fine-tuning
problem for the lightest CP-even Higgs boson mass in the MSSM, and
generated the observed baryon asymmetry via electroweak
baryogenesis. In addition, we calculate the Higgs boson masses,
chargino masses and neutralino masses at tree level, and predict the
new Higgs physics which can be tested at the LHC and ILC.
Confirmation one of these models at the future colliders might give
us indirect implication of string theory.

{\bf Acknowledgments~--}~This research was supported in part by the
Cambridge-Mitchell Collaboration in Theoretical Cosmology.


\end{document}